\documentclass[journal]{IEEEtran}

\usepackage{xcolor,soul,framed} 

\colorlet{shadecolor}{yellow}
\usepackage[pdftex]{graphicx}
\graphicspath{{../pdf/}{../jpeg/}}
\DeclareGraphicsExtensions{.pdf,.jpeg,.png}

\usepackage[cmex10]{amsmath}
\usepackage{array}
\usepackage{mdwmath}
\usepackage{mdwtab}
\usepackage{eqparbox}
\usepackage{url}

\usepackage[caption=true,font=normalsize,labelfont=sf,textfont=sf]{subfig}
\usepackage{stfloats}
\usepackage{algorithmic}
\usepackage{algorithm}
\begin{document}

\bstctlcite{IEEEexample:BSTcontrol}
    \title{Cross: A Delay Based Congestion Control Method for RTP Media}
  \author{Songyang Zhang,~\IEEEmembership{Student Member,~IEEE,}
      Changpeng Yang,~\IEEEmembership{Student Member,~IEEE,}

  \thanks{ Songyang Zhang(e-mail: sonyang.chang@foxmail.com). I am no longer working on congestion control. I post it to arxiv since there are some interesting results.}
  }

\markboth{Journal of \LaTeX\ Class Files,~Vol.~14, No.~8, August~2024
}{Roberg \MakeLowercase{\textit{et al.}}: High-Efficiency Diode and Transistor Rectifiers}

\maketitle

\begin{abstract}
After more than a decade of development, real time communication (RTC) for video telephony has made significantly progress. However, emerging high-quality RTC applications with high definition and high frame rate requires sufficient bandwidth. The default congestion control mechanism specifically tuned for video telephony leaves plenty of room for optimization under high-rate scenarios. It is necessary to develop new rate control solutions to utilize bandwidth efficiently and to provide better experience for such services. A delay-based congestion control method called Cross is proposed, which regulates rate based on queue load with a multiplicative increase and multiplicative decrease fashion. A simulation module is developed to validate the effectiveness of these congestion control algorithms for RTC services. The module is released with the hope to provide convenience for RTC research community. Simulation results demonstrate that Cross can achieve low queuing delay and maintain high channel utilization under random loss environments. Online deployment shows that Cross can reduce the video freezing ratio by up to 58.45\% on average when compared with a benchmark algorithm. 
\end{abstract}


%
\IEEEpeerreviewmaketitle


\section{Introduction}

\IEEEPARstart{I}{n} May 2011, Google released an open source project WebRTC to enable real-time communication through browser. It is widely adopted by applications to build video telephony and conference service, enabling billions of users to establish face-to-face communication \cite{webrtc-market}. 

To ensure a high-quality experience, real-time communication (RTC) applications require low latency and sufficient bandwidth. The best effort delivery mode of the Internet has no assurance on service quality. Fluctuating bandwidth, large latency, and lost packets can lead to stuttering and mosaic, which are harmful to user engagement \cite{Krishnan2012}. Therefore, much effort has been devoted by both industry and academia to provide better service through data transmission optimization. This includes integrating forward error correction (FEC)\cite{rfc8854}, upgrading congestion control algorithms (CCAs), and steering traffic in overlay networks\cite{Jiang2016}.

Congestion control algorithm is a crucial component in data transmission protocol. According to the network status, it constantly adjusts the data transmission rate to cooperatively share bandwidth. Traditional congestion control algorithms in Transmission Control Protocol (TCP) are maximum throughput oriented. Most of them continuously increase rate until buffer overflows in routers and high transmission delay is thus introduced. To provide satisfactory Quality of Experience (QoE), Google Congestion Control (GCC) is deployed in WebRTC. GCC mainly targets for video telephony service, and a maximum bitrate of 2.5Mbps is sufficient. WebRTC is now used to create immersive real-time video services, including cloud gaming, virtual tours, and online museums. They usually rely on remote cloud servers that have high-end graphics processing units (GPUs) to generate eye-catching content. These high-quality interactive video streaming requires more bandwidth, typically ranging from 10-35Mbps \cite{cloud-gaming}. The bandwidth increment in GCC is sluggish for high-quality RTC.

Hence, it is necessary to develop new rate control methods for high-quality RTC services. Given the excellent performance of TCP BBR\cite{Cardwell2016}, many developers try to implant BBR on WebRTC with hope to achieve better performance. Unfortunately, it is found that BBR has deteriorated performance in WebRTC. In this work, we propose a new congestion control algorithm called Cross, which mainly targets for high-quality RTC services. According to the number of buffered packets in bottleneck, Cross adjusts the target rate with a multiplicative increase and multiplicative decrease (MIMD) fashion. Cross achieves low transmission delay by quickly responding to queue accumulation. To improve video playback smoothness when link capacity is suddenly dropped, a subroutine to quickly adapt to bandwidth is implemented. 

A simulation module is developed in this work to run WebRTC on ns-3 \cite{ns3}. In a fully controllable environment, the performance of different congestion control algorithms can be evaluated and each test is reproducible. It is convenient for researchers to locate lurking pitfalls and to find potential improvements for newly designed congestion control algorithms. The code of simulation module can be accessed at \cite{webrtc-gcc-ns3}. And it has already been integrated into Opennetlab Gym \cite{Eo2022}, which is released by Microsoft Research Asia to enable researchers to explore artificial intelligence (AI) driven congestion control solutions for RTC services.

The key contributions of this paper are summarized as follows:
\begin{itemize}
    \item A congestion control algorithm targeting for high-quality real-time video streaming is proposed. By discreetly maintaining a constant queue load at the bottleneck, Cross can reach a fair bandwidth allocation and is highly resistant to random packet loss. By deploying on an online service, results demonstrate that Cross can achieve better performance in term of video playback smoothness. 
    \item Under simulated environment, we observed that WebRTC BBR experiences excessively high packet loss rate. We analyze the reason why BBR cannot collaborate well with WebRTC and potential solutions for better performance is provided.
    \item  Evaluation results on several other algorithms are presented for interested readers. We hope that the simulation module can provide some convenience for the development of congestion control algorithms for RTC services.
\end{itemize}

The remainder of this paper is organized as follows. Section \uppercase\expandafter{\romannumeral2} first introduces the relevant works on congestion control. Section \uppercase\expandafter{\romannumeral3} provides a detailed description of the simulation module. Section \uppercase\expandafter{\romannumeral4} presents the diagnosis on BBR and potential workaround. Section \uppercase\expandafter{\romannumeral5} gives the design details of Cross. Section \uppercase\expandafter{\romannumeral6} presents the evaluation results on Cross and several benchmark algorithms. Section \uppercase\expandafter{\romannumeral7} concludes this article.
\section{Releated work}
In 1986, the first congestion collapse event happens and throughput decreased by three orders of magnitude. Jocobson recommended the additive increase and multiplicative decrease (AIMD) law to regulate the transmission rate of TCP \cite{Jacobson1988}. According to whether there is loss packet or not, AIMD adjusts the congestion window size (CWND) to limit the in-flight packets. By implementing congestion control algorithm at end hosts, the network resource is allocated in distributed manner. With AIMD, different flows within a group can eventually converge to similar throughput to achieve bandwidth fairness \cite{Chiu1989}, which is a sufficient condition for network stability. AIMD saved the Internet from congestion collapse and laid a solid foundation for most congestion control algorithms developed later. 

AIMD is far from perfect since its inception. In high bandwidth-delay product (BDP) networks, AIMD takes a long time to reach the maximum throughput again after a single packet drop. BIC \cite{Xu2004}, CUBIC \cite{Ha2008} and Elastic \cite{Alrshah2019} are proposed to address this issue. Cubic takes a cubic function to search for the upper limit on CWND while Elastic defines a window-correlated weight function for acceleration. In wireless networks, random packet loss can occur due to noise, fading, mobility, and interference \cite{Gurtov2004}. Reducing CWND under such circumstances has a negative impact on throughput and leaves the channel in idle status for most of time. To enhance throughput, Westwood \cite{Casetti2002} resets CWND based on estimated bandwidth-delay product.

Some other congestion control algorithms, such as Vegas \cite{Brakmo1995}, Fast \cite{Wei2006}, and Copa \cite{Arun2018a}, adjust CWND based on delay signal. By quickly reacting to queue accumulation, these protocols can achieve low transmission delay. However, when competing with loss-based congestion control algorithms, these delay-based protocols will get starved. In an era where most applications target maximum throughput, delay-based congestion control algorithms are not widely used. What’s more, these delay-based solutions map the sending rate to queuing delay and they must map a wide range of rates into a small delay range to work across different rates \cite{Arun2022}. Hence, non-congestive delay jitter will induce enormous rate fluctuation.

With declining memory prices and the fallacy that more is better, intermediate routers are provisioned with large buffer. Loss-based congestion control algorithms has lead Bufferbloat \cite{Gettys2011}. The increased latency becomes unbearable for ever-growing delay sensitive traffic. It is a consensus to design congestion control method which could simultaneously achieve high throughput and low delay. Several solutions have been proposed, e.g. PCC\cite{Dong2015}, Vivace \cite{Dong2018}, BBR and Copa. Among them, BBR stands out as the most eminent one and has caused a stir in both academia and industry. Mathis \cite{Mathis2019} claims that BBR marks the beginning of a new era in congestion control and represents a radical departure from past solutions\cite{Mathis1997}. According to a utility function, PCC and Vivace continuously adjust rate in an online fashion. However, it is a challenge to select proper coefficients for each term in the utility function with different units under different circumstances.

To promote bandwidth fairness and to improve QoE, congestion control is deployed in RTC services. The RTP Media Congestion Avoidance Techniques (RMCAT) working group was formed to develop congestion control solutions for interactive real-time traffic. The group has developed three algorithms: GCC, NADA \cite{rfc8698} and SCReAM \cite{rfc8298}. To reduce self-inflicted queue latency, all three protocols rely on delay related signal to adjust the data transmission rate. GCC uses delay gradient to detect channel overuse. NADA adjusts the target rate of video encoder according to aggregated delay and SCReAM uses queue delay. Reference \cite{Addanki2022} indicates the equilibrium point for delay gradient based CCAs is not unique the bandwidth fairness cannot be guaranteed.

Some recent works \cite{Jiang2021, Yen2023} have attempted to use deep reinforcement learning methods to model congestion control as a sequential decision-making process. These methods rely on a neural network that maps network states to congestion control actions. In simulated environments, it is claimed by their authors that these black-box solutions demonstrate better performance than traditional model-based solutions. However, most of them have not been tested in wild networks and many developers are hesitant to deploy them in their applications due to the black-box nature.
\begin{figure*}
\centering
\includegraphics[width=7in]{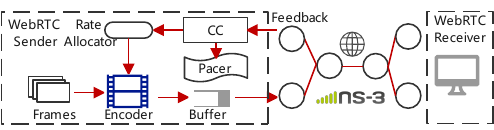}
\captionsetup{justification=centering}
\caption{The simulation module to run WebRTC on ns-3}
\label{Fig:framework}
\end{figure*}

\section{Overview of simulation platform}
A module called ``ex-webrtc" is developed in ns-3 to run WebRTC natively. As illustrated in Figure \ref{Fig:framework}, the simulation tool consists of three main components: the WebRTC sender, the WebRTC receiver, and the network environment. 

At the sender side, the following components are included: frame generator, video encoder, congestion controller (CC), rate allocator, pacer. During simulation, the frame generator creates raw frames at fixed intervals based on frame rate. For simplicity, the pixels of each frame are padded with random values. Reading data from YUV sequences is also supported. For frames with random values, the input data is ignored and the simulated encoder will output data based on target rate. For YUV frames with valid data, a real video codec (h264 or vp8) should be applied for compression. Then compressed image is packetized into RTP packets. The pacer can evenly send packets into network and its rate is modulated by congestion controller. The packets are then sent into the ns-3 network environment through socket interfaces. From RTCP transport feedback message, packet loss and one-way transmission delay can be extracted. Such information is fed into the congestion controller to adjust target rate.

The receiver is responsible for parsing RTP packets, reassembling encoded images, decoding video frames, rendering content and sending feedback message.

In ns-3, a node can act as a host with full TCP/IP stack. With several nodes connected by point-to-point links, we could build a small-scale network. We could set bandwidth, queue discipline buffer length and propagation delay in link. The simulator continuously schedules these queued discrete events. Packet will be processed and delivered to its destination. The asynchronous events and delayed tasks in WebRTC are handled by several task queues and these task callbacks are processed by ns-3 scheduler. An external clock is registered in WebRTC to get simulation time.

Once a network topology is built, multiple WebRTC flows can be installed on nodes and each flow can choose a specific congestion control algorithm. With multiple flows contending for bandwidth, the fairness and stability of a specific algorithm can be evaluated. Additionally, to evaluate inter-protocol friendliness, these congestion control algorithms (Reno, Cubic, BBR) in TCP can be running in parallel by installing bulk data transfer application. 

Metrics are collected into trace files to analyze the performance of different congestion control algorithms. Before calling socket send function, a tag containing the sequence number and sent timestamp is piggybacked in the ns-3 packet object. At the sender side, the target rate is output at regular intervals to capture the rate dynamic of a session. When a media packet arrives at its destination, the one-way transmission delay (OWD) is calculated and the length of the packet is counted. OWD indicates buffer occupation status in routers. Every second, the receiver computes goodput. When each session comes to an end, the packet loss rate and average owe-way transmission delay is calculated.
\section{Investigation on Webrtc BBR}
\begin{figure}
\includegraphics[height=0.8in, width=2.5in]{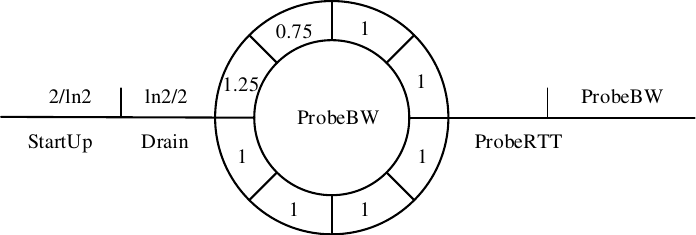}
\centering
\caption{BBR state transition diagram}
\label{Fig:state}
\end{figure}

In an old version \cite{webrtc-bbr}, the implementation on BBR is included in WebRTC. We obtain the code running on the simulation module and analyze its performance.

BBR determines the number of in-flight packets with two control parameters: the pacing rate and CWND. The pacing rate determines how fast the packets are injected into network, and CWND sets an upper limit on the number of in-flight packets. CWND is determined by the product of the estimated bandwidth and the minimum roundtrip time ($RTT_{min}$). When a sent packet is acknowledged, a sample on roundtrip time and the length of the delivered packet ($\Delta delivered$) during this round can be obtained. A bandwidth sample is thus be calculated as Equation (\ref{eq:bw-sample}). The estimated bandwidth ($bw_{es}$) is got from a maximum window filter with length of 10 rounds. The pacing rate in Equation (\ref{eq:pacing-rate}) is the product of pacing gain and estimated bandwidth.
\begin{equation}
\label{eq:bw-sample}
bw=\frac{\Delta delivered}{\Delta t}
\end{equation}
\begin{equation}
\label{eq:pacing-rate}
pacing\_rate= bw_{es} \times pacing\_gain
\end{equation}

As illustrated in Figure \ref{Fig:state}, BBR operates in four states: StartUp, Drain, ProbeBW, and ProbeRTT. 

In StartUp state, the delivery rate of BBR increases exponentially by a gain of 2/ln2 in each round to probe maximum available throughput. When the estimated bandwidth in current round is less than the target rate ($1.25\times bw_{es}$) for three rounds, the state is switched to Drain. To free up excess buffered packets during StartUp, BBR slows down pacing rate by ln2/2 gain in Drain until the in-flight packets match with BDP. 

There are eight cycles in ProbeBW and each cycle lasts for at least one $RTT_{min}$. In probe up cycle, the pacing gain is 1.25 to probe extra bandwidth. In probe down cycle, the pacing gain is 0.75 to get rid of possible queue accumulation in previous cycle. The next six rounds are the probe cruise cycles with 1.0 as pacing gain. CWND is equal to $2\times bw_{es}\times RTT_{min}$ in ProbeBW state.

If the minimum RTT is not sampled again within 10 seconds, BBR assumes that the network is in congestion. It then enters the ProbeRTT state and CWND is reduced to 4 packets to clear buffered packets in network. Additionally, $RTT_{min}$ is updated with newly measured values. During steady state, BBR switches between ProbeBW and ProbeRTT.

WebRTC utilizes the estimated bandwidth in congestion controller to configure the target rate of the video encoder. When multiple flows contend for resource, BBR tend to overestimate bandwidth and persistent queue still forms in bottleneck\cite{Hock2017}. Transmission delay increment is unavoidable. During ProbeBW state, when RTT is larger than $2* RTT_{min}$, $bw_{es}* RTT > CWND = 2*bw_{es}* RTT_{min}$ can be deduced. If the length of in-flight packets is larger than CWND, the packet sending task will be temporarily blocked. No packet can be sent out until a feedback is received. However, the encoder continues producing encoded images and packets will be accumulated in the pacer queue, which leads obvious video lag. Same issue is also reported in \cite{bbr-question}. The interval to send feedback messages in WebRTC is between 50 and 250 milliseconds, which further exacerbates this problem. To ensure low latency, the pacing controller recalculates the sending rate when queued packets cannot be drained within a predefined threshold (queue\_time\_limit\_). In such scenario, packet sending rate is out of control of BBR. If the key of drain\_large\_queue\_ is tuned to false, rate recalculation will not happen but unbearable latency will be induced when there are too many packets in the pacer queue.

One intuitive solution is to remove the CWND limit. Since the bandwidth overestimation nature of the max window filter, the queue occupancy may grow indefinitely \cite{Arun2022}. CWND plays a significant role to cap the number of in-flight packets to avoid excess packet loss.We propose Rate BBR (RBBR), which removes the CWND limit and introduces two other parameters: pacing discount and target discount in the ProbeBW state. By applying different discount values on the pacing rate, RBBR limits the number of in-flight packets in the network pipe. The target discount parameter reduces the bitrate of the encoder to avoid excess packets buffered at the pacer queue. This design is to prevent rate recalculation. Algorithm \ref{alg:discount} details how RBBR updates the target rate during ProbeBW state. Further, RBBR cuts the rate by half in the ProbeRTT state to reduce packet accumulation in the pacer queue. And the expiry time for minimum RTT is 5 seconds. To maintain the target rate stability, the values on probe up gain and probe down gain are 1.125 and 0.875. And the cycle length is 4 in RBBR. If extra bandwidth is available, the rate will increase by 1.27 (1.125*1.125) after 8 cycles.

\begin{algorithm}[H]
\caption{CreateRateUpdate}\label{alg:discount}
\begin{algorithmic}[1]
\REQUIRE ~~\\ 
$inflight$, $pacer\_queue$
\ENSURE ~~\\ 
$target\_rate$, $pacing\_rate$
\STATE {$bw \gets BandwidthEstimate()$}
\STATE {$target\_discount \gets 1.0$}
\STATE {$pacing\_discount \gets 1.0$}
\STATE {$bdp \gets BandwidthEstimate() * GetMinRtt()$}
\STATE{$target\_rate \gets bw * pacing\_gain\_$}
\STATE{$pacing\_rate \gets bw * pacing\_gain\_$}
\STATE {$expect \gets \frac{ pacer\_queue}{pacing\_rate}$}
\IF{$expect \ge GetMinRtt()$}
	\IF{$0 == cycle\_current\_offset\_$}
		\STATE{$target\_rate \gets bw$}
	\ENDIF

	\IF{$expect \leq 2.0 * GetMinRtt()$}
		\STATE{$target\_discount \gets 0.95$}
	\ELSIF{$expect \leq 2.5 * GetMinRtt()$}
		\STATE{$target\_discount \gets 0.90$}
	\ELSIF{$expect \leq 3.0 * GetMinRtt()$}
		\STATE{$target\_discount \gets 0.85$}
	\ELSE
		\STATE{$target\_discount \gets 0.80$}
	\ENDIF
	\STATE{$target\_rate \gets bw * target\_discount$}
\ENDIF

\IF{$inflight\neq \infty\AND\ cycle\_current\_offset\_ \geq 1$}
	\IF{$inflight \ge 3.0*bdp$}
		\STATE {$pacing\_discount \gets 0.7$}
	\ELSIF{$inflight \ge 2.5*bdp$}
		\STATE {$pacing\_discount \gets 0.8$}
	\ELSIF{$inflight \ge 2.0*bdp$}
		\STATE {$pacing\_discount \gets 0.9$}
	\ELSIF{$inflight \ge 1.5*bdp$}
		\STATE {$pacing\_discount \gets 0.95$}
	\ENDIF
	\STATE{$rate \gets pacing\_discount * bw$}
	\STATE{$target\_rate \gets min(target\_rate,rate)$}
	\STATE{$pacing\_rate \gets min(pacing\_rate, rate)$}
\ENDIF
\end{algorithmic}
\label{discount}
\end{algorithm}

\begin{figure}[!htb]
\subfloat[BBR]{
\begin{minipage}[t]{0.5\linewidth}
    \includegraphics[width = 1.7in]{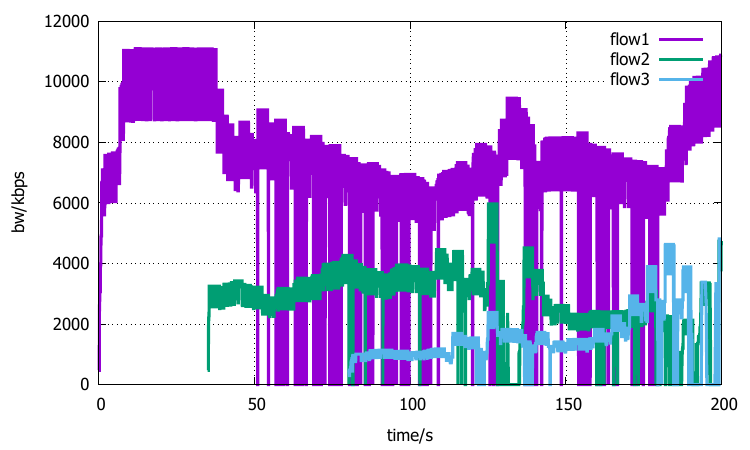}
\end{minipage}}
\subfloat[RBBR]{
\begin{minipage}[t]{0.5\linewidth}
    \includegraphics[width = 1.7in]{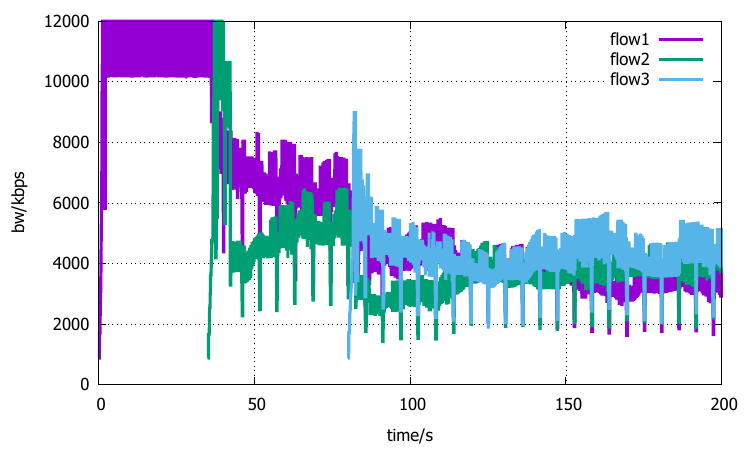}  
\end{minipage}}
\centering
\caption{Target transfer rate of flows in Case2.}
\label{Fig:bbr-rbbr} 
\end{figure}

A point-to-point network is built to compare the performance of BBR and RBBR. The parameters in Table \ref{tab:p2pcfg} are bandwidth (in unit of Mbps), one-way propagation delay (in unit of milliseconds) and queue length in net device. Results on target transfer rate of BBR and RBBR flows are given in Figure \ref{Fig:bbr-rbbr}. Due to packets piling up in the pacer queue, rate recalculation frequently occurs in WebRTC-BBR. If the expected queue time exceeds 2 seconds, WebRTC temporarily pauses the encoder and sets the bitrate to zero. These zero values can be observed in Figure \ref{Fig:bbr-rbbr}(a), and the first flow achieves the highest rate. RBBR flows converge to bandwidth fairness in Case2, as illustrated in Figure \ref{Fig:bbr-rbbr}(b). By removing the CWND limit in the ProbeBW state, we evaluate BBR in QUIC (denoted as QUIC-BBR-NCWND) on ns-3. As shown in Table \ref{tab:avgloss}, QUICBBR-NCWND has the highest packet loss rate. Compared with WebRTC-BBR, RBBR has lower packet loss rate in all three tests. 

In simulation platform, the capacity of the bottleneck is small and several flows can saturate it. In a real network, WebRTC-BBR may provide enough bandwidth and the output bitrate of video encoder may be below the available bandwidth. In such scenario, rate recalculation rarely happens. To avoid abnormality and prevent excess packet loss, it is indeed a responsible behavior for the WebRTC team to remove BBR from the codebase.

\begin{table}[!t]
\caption{Link Configuration}
\centering
\label{tab:p2pcfg}
\begin{tabular}{|c|c|c|c|}
\hline
Case & Bandwidth & OWD & Queue Length \\ \hline
1    & 6Mbps     & 100ms            & 6Mbps*300ms  \\ \hline
2    & 12Mbps    & 50ms             & 12Mbps*150ms \\ \hline
3    & 15Mbps    & 30ms             & 15Mbps*100ms \\ \hline
\end{tabular}
\end{table}
\begin{table}[!t]
\caption{Packet Loss Rate}
\centering
\label{tab:avgloss}
\begin{tabular}{|c|c|c|c|}
\hline
Loss Rate(\%)         & Case1 & Case2 & Case3 \\ \hline
WebRTC-BBR  & 3.49  & 5.90  & 3.89  \\ \hline
WebRTC-RBBR & 0.62  & 1.99  & 0.25  \\ \hline
QUIC-BBR    & 2.06  & 1.21  & 0.47  \\ \hline
QUIC-BBR-NCWND & 8.34  & 7.87  & 8.83  \\ \hline
\end{tabular}
\end{table}

\section{Design of Cross}
Rate control solution for RTC service should adapt video encoder to efficiently utilize channel resource while maintaining low delay. For high-quality RTC streaming, the way to linearly increase bandwidth in GCC is slowly. BBR probes bandwidth in MIMD fashion, but its available bandwidth is estimated from acknowledged packets.  The output bitrate of the encoder may not match with the target rate specified by the congestion controller. When the video frames are stationary, the encoder usually requires less bitrate. For motion content, the encoder generates more bitrate. When fewer packets are injected into network, updating bandwidth filter frequently with recent rate samples will under-estimate bandwidth. If bandwidth filter is not updated timely in application limit mode, video display may freeze when link capacity drops and many packets are buffered in the bottleneck. The issue of bandwidth under-estimation can be addressed by sending padding packets, but this is a waste of bandwidth, especially for high-quality real-time video streaming.
\begin{equation}
\label{eq:queue-load}
B = x(t)*(D_{m}-D_{min})
\end{equation}
\begin{equation}
\label{eq:mimd}
x(t+1)=
\begin{cases}
x(t)*(1+\eta),& \text{if}\ B \leq B_{th}\\
x(t)*(1-\eta),& \text{otherwise}
\end{cases}
\end{equation}
\begin{figure}
\includegraphics[height=1in, width=3in]{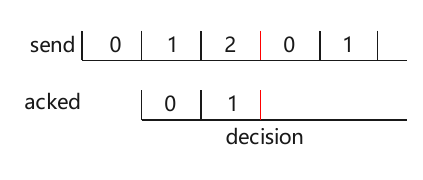}
\centering
\caption{Monitor cycles}
\label{Fig:mi}
\end{figure}

Cross is designed for high-quality real-time video applications. Cross probes bandwidth at fast pace and tries to maintain queue occupation at a low level. Cross uses queue load defined in Equation \ref{eq:queue-load} to infer whether the network pipe falls into congestion. $x(t)$ is current rate and $B$ denotes the extra packets buffered in network. $D_m$ measures the bloated delay due to congestion and $D_{min}$ is the minimum delay within the monitor time window. According to whether the extra buffered packets exceed the defined threshold ($B_{th}$), Cross adjusts target rate in MIMD way as given in Equation \ref{eq:mimd}. $B_{th}$ is a tunable factor to control the number of packets buffered at the bottleneck. It should be noted that Cross draws much inspiration from Vegas, Copa, PCC, and BBR.

In WebRTC, Cross uses one-way delay samples from feedback messages to measure bloated delay $D_{m}$. Time is divided into cycles and each cycle lasts about one $RTT_{min}$. Each cycle is indexed by cycle\_current\_offset\_, same notation from BBR. The start time of each cycle will be recorded and Cross advances offset in OnSentPacket function when duration exceeds $RTT_{min}$. The duration for each cycle is more accurate if it were solely advanced by the feedback arrival events. For each decision epoch, the sequence and sending timestamp of these packets sent at cycle 0 and cycle 1 are stored. As illustrated in Figure \ref{Fig:mi}, We assume the acknowledgement of a specific packet is delayed for at least one $RTT$. When the packet with the largest sequence number in cycle 1 is acknowledged, the cycle offset will be reset to zero and a new decision epoch is started. The one-way delay samples in $owd\_array$ in cycle 1 are sorted and the median delay $OWD_{m}$ defined in Equation \ref{eq:owd-median} is used for decision-making. The number of elements in $owd\_array$ is $n$. $numerator$ and $denominator$ are two tunable factors. By feeding $OWD_{m}$ and $OWD_{min}$ into Equation \ref{eq:queue-load}, Cross decides whether to move the rate to UP direction or DOWN direction. $OWD_{m}- OWD_{min}$ measures queue delay and time synchronization on both sides is unnecessary.  
\begin{equation}
\label{eq:owd-median}
OWD_{m} = owd\_array[\frac{n * numerator}{denominator}]
\end{equation}

The property on bandwidth allocation fairness of Cross is analyzed here. Assuming there are two flows A and B traversing the same bottleneck and their transmission rates are $x_a(t)$ and $x_b(t)$. What’s more, $x_a(t)$ is larger than $x_b(t)$. When the queue load $B$ is below the threshold, both flows will increase their rates. The ProbeRTT state in Cross assures flows can sample the minimum value for $OWD_{min}$ when buffer is empty in the bottleneck. Since the values on queue delay sampled by both flows are quite close, flow A will first cross queue load threshold $B_{th}$. When such event happens, flow A has to reduce bandwidth and B can increase its rate further. As time goes by, the rate difference between two flows decreases and their rates move toward fairness line. To penalize flows with large rate when network falls into congestion can promote fairness.
\begin{algorithm}[H]
\caption{RateAdjustmentInProbeBW}\label{alg:cross}
\begin{algorithmic}[1]
\REQUIRE ~~\\ 
$act$, $OWD_{m}, OWD_{min}$
\STATE {$gain\_up \gets 1 + \eta$}
\STATE {$gain\_down \gets 1- \eta$}
\IF{$DOWN == act\ \AND\ UP == last\_action\_$}
\STATE {$max\_rate\_ \gets cur\_rate\_$}
\STATE {$gain\_down \gets \epsilon$}
\ENDIF
\IF{$UP == act\ \AND\ act == last\_action\_$}
\STATE {$num\_up\_\gets num\_up\_ +1$}
\ELSE
\STATE {$num\_up\_\gets 0$}
\ENDIF
\IF{$DOWN == act\ \AND\ act == last\_action\_$}
\STATE {$num\_down\_\gets num\_down\_ +1$}
\ELSE
\STATE {$num\_down\_\gets 0$}
\ENDIF
\STATE {$ AR\gets 0$}
\IF{$num\_up\_ > k_{up}$}
	\STATE {$ T\gets limit$}
	\IF{$max\_rate\_ \neq 0$}
		\IF{$max\_rate\_ \ge cur\_rate\_$}
			\STATE {$ T\gets 0.5*(max\_rate\_ - cur\_rate\_)$}
			
		\ELSE
			\STATE {$ T\gets cur\_rate\_ - max\_rate\_$}
		\ENDIF
	\ENDIF
	\STATE {$ AR\gets min(T,limit)$}
\ENDIF
\IF{$num\_down\_> k_{down}$}
	\STATE {$gain\_down \gets d$}
	\STATE {$num\_down\_\gets 0$}
\ENDIF
\IF{$UP == act$}
	\STATE {$ D_{th} \gets OWD_{min} + k * GetMinRtt()$}
	\IF{$OWD_m \leq D_{th}$}
		\STATE {$cur\_rate\_\gets gain\_up *(cur\_rate\_ + AR)$}
	\ELSE
		\STATE {$cur\_rate\_\gets gain\_up *cur\_rate\_$}
	\ENDIF
\ELSIF{$DOWN == act$}
	\STATE {$cur\_rate\_\gets gain\_down *cur\_rate\_$}
\ENDIF
\STATE {$last\_action\_ \gets act$}
\end{algorithmic}
\end{algorithm}

Cross inherits the four states present in BBR. In the StartUp state, Cross make a fast probe to the maximum available bandwidth in a manner similar to BBR. Additionally, when the queue load $B$ exceeds $B_{th}$ for three consecutive rounds, Cross enters into Drain state. In the ProbeBW state, Cross regulates its rate based on the rule in Equation \ref{eq:mimd}. If $RTT_{min}$ is not sampled again within 5 seconds, Cross enters into the ProbeRTT state. In this state, the rate is cut by half to sample new values for $RTT_{min}$ and $OWD_{min}$. By periodically draining the buffered packets from the bottleneck link, flows entering into network at different time have the opportunity to sample the minimum delay. If flows traversing the same bottleneck measure different congestion prices, it is difficult for delay-based solutions to achieve bandwidth allocation fairness. Since different flows sample different values on base RTT, it is well known that Vegas has later comer advantage effect \cite{AlSaadi2019}: the new flows achieve higher throughput than pre-existing flows. The results in evaluation part demonstrate NADA suffers from the same issue.

The detail of rate adjustment is given in Algorithm \ref{alg:cross}. Similar to PCC, Cross increases or decreases its rate by $\eta$ (0.05) at each decision time point. As illustrated by Figure \ref{Fig:mi}, Cross makes rate decision for about every 3 rounds. The bandwidth increment process of Cross appears to be slow in a network with fluctuating capacity. For a path with 50ms roundtrip delay, it takes about 750ms to reach 2.55Mbps from 2.0Mbps. A higher rate adjust factor e.g. $\eta= 0.1$ tends to reduce channel utilization. An additional rule is introduced for rate acceleration if delay $OWD_{m}\leq D_{th}$ is at low level (line 35, k = 0.125). When the number of times that Cross is in UP direction exceeds $k_{up}$, additional rate factor is calculated (line 18-27) and the upper limit (200kbps) is applied to prevent overshooting too much. When the direction is switched from UP to DOWN, the link is in congestion and the last rate increment decision is a failure. Hence, a lower reduce factor $\epsilon =(1-\eta)/(1+\eta) $ is used to slightly drain buffered packets from bottleneck (line 5). The factor $1/(1+\eta)$ is to recover the rate before and $(1-\eta)$ is to reduce rate further. Correspondingly, a fast reduce gain $d = 0.85$ (line 30) is introduced to yield more bandwidth when Cross moves toward DOWN direction for more than $k_{down}$ times. Experiments demonstrated that it helps rate convergence when new flows come.

In real-world networks, data flows may experience sudden bandwidth drop. According to tests conducted by Tencent Start, approximate 5.6\% of Ethernet users and 35.5\% of WiFi users encounter at least five occurrences of significant bandwidth reduction (reduction by more than 50\% within 100 ms) per minutes. Sudden bandwidth reduction can occur due to background traffic surges. In wireless networks, signal collisions and user movements can also cause significant fluctuations in link capacity. In wired networks with a shared queue, rapidly decreasing the rate of a single flow may not effectively alleviate congestion. However, for wireless access, the last hop is often the bottleneck. As indicated in \cite{Abbasloo2019}, the base station is equipped with per-UE queue for traffic isolation and to ensure user-level fairness. Smart Queue Management such as fq\_codel \cite{rfc8290} in home routers, performs per-flow scheduling to mitigate bufferbloat. Overall, the ability to rapidly adapt to changes in available bandwidth can reduce self-inflicted queue delay and improve video playback smoothness. 

Cross adjusts its rate with small amplitude. To quickly respond to a sudden bandwidth drop and to reduce buffered packets, a fast bandwidth adaptation method is implemented in Cross and Algorithm \ref{alg:bw-detection} provides the details. Since I-frame is larger than P-frame, the peak rate is usually higher than the average rate \cite{codec-mode}. To accommodate such burst traffic, it is common to set an upper bandwidth in congestion controller higher than the expected maximum average bitrate of encoder. In a short time period, the output bitrate may be smaller than current rate in congestion controller. Using the cur\_rate and acknowledged rate for bandwidth change detection will lead error decision. When a feedback message arrives, multiple samples of sending rate and acknowledged rate can be obtained with the help of BandwidthSampler, which is a functionality implemented in BBR. If the maximum acknowledged rate ($ack\_rate_{m}$) within one RTT is less than the maximum sending rate ($send\_rate_{m}$) by a factor $ratio=0.75$ (line 5), Cross deems that a sudden bandwidth reduction event occurs. When $ ack\_rate_{m}$ is larger than low bandwidth level ($bw\_lo$), current rate is reset as $\beta * ack\_rate_{m}$ ($\beta= 0.75$). Otherwise, current rate is assigned with the value of $\beta * max(real\_rate, ack\_rate_{m})$. This is the case where the sent packets are not enough and $ack\_rate_{m}$ is too small. $bw\_lo$ is the minimum rate specified by applications. $good\_put\_$ is a window filter and it samples the maximum delivered rate within 5 seconds.
\begin{algorithm}[H]
\caption{FastBandwidthAdaptation}\label{alg:bw-detection}
\begin{algorithmic}[1]
\REQUIRE ~~\\ 
$send\_rate_{m}$, $ack\_rate_{m}$,  $standing\_owd$
\ENSURE ~~\\ 
$enter\_probe\_rtt$
\STATE {$enter\_probe\_rtt \gets 0$}
\STATE {$target \gets OWD_{min} + f*GetMinRtt()$}
\STATE {$excess \gets OWD_{min} + m*GetMinRtt()$}
\IF{$standing\_owd > target\ \AND\ ack\_rate_{m}\neq 0$}
	\IF{$ratio * send\_rate_{m} > ack\_rate_{m}$}
		\STATE {$real\_rate \gets min(cur\_rate\_,good\_put\_.GetBest())$}
		\IF{$ack\_rate_{m} >bw\_lo$}
			\IF{$standing\_owd > excess$}
				\STATE {$cur\_rate\_\gets ack\_rate_{m}$}
				\STATE {$enter\_probe\_rtt \gets 1$}
			\ELSE
				\STATE {$cur\_rate\_\gets \beta * ack\_rate_{m}$}
			\ENDIF
		\ELSE
			\STATE {$cur\_rate\_\gets \beta * max(real\_rate,ack\_rate_{m})$}
		\ENDIF
	\ENDIF
\ENDIF
\end{algorithmic}
\end{algorithm}

$standing\_owd$ is the minimum one-way delay obtained from standing window filter, which is defined by Copa. And its implementation can be referred in mvfst \cite{mvfst}. When delay is bloated to some extent ($standing\_owd >target$, line 4, f = 0.25), this subroutine takes effect. When delay is excessively high (line 8-10, m=2.0), Cross resets bandwidth and enters ProbeRTT. Such action is to let buffered packets drained from the bottleneck. Delay threshold is introduced to enhance robustness. When channel capacity deteriorates or background traffic surges, it is a fact that flow can observe increased delay. Once sudden bandwidth reduction happens and the estimated bandwdith is reset, this algorithm will not be triggered again for the next two rounds.
\section{Evaluation}
First, we evaluate the performance of these congestion control algorithms on a point-to-point link, including Cross, GCC, Vivace, Copa, RBBR, NADA, and NADAR. Then we deploy Cross in an online service and compare its performance in term of video freezing duration ratio with BBR. The code for GCC and Vivace is available in WebRTC. We also implement NADA and Copa in WebRTC. In NADA, we introduce the ProbeRTT state and rename it as NADAR for distinction. In Cross, the queue load threshold $B_{th}$ is equal to 2*MTU (Maximum Transmission Unit). The parameters on numerator and denominator in Equation \ref{eq:owd-median} are 1 and 4 in respectively. The value for $k_{up}$ and $k_{down}$ is 3.
\begin{table}[]
\caption{Link Configuration}
\centering
\label{tab:link-cfg}
\begin{tabular}{|c|c|c|c|}
\hline
Case & BW(Mbps) & OWD(ms) & Queue Length \\ \hline
1    & 3        & 30      & 3Mbps*90ms   \\ \hline
2    & 3        & 50      & 3Mbps*200ms  \\ \hline
3    & 6        & 30      & 6Mbps*90ms   \\ \hline
4    & 6        & 50      & 6Mbps*200ms  \\ \hline
5    & 12       & 30      & 12Mbps*90ms  \\ \hline
6    & 12       & 50      & 12Mbps*200ms \\ \hline
7    & 15       & 30      & 15Mbps*90ms  \\ \hline
8    & 15       & 50      & 15Mbps*200ms \\ \hline
\end{tabular}
\end{table}

\begin{figure}[htbp]
\centering
\begin{minipage}[t]{0.49\linewidth}
\centering
\includegraphics[width=1\linewidth]{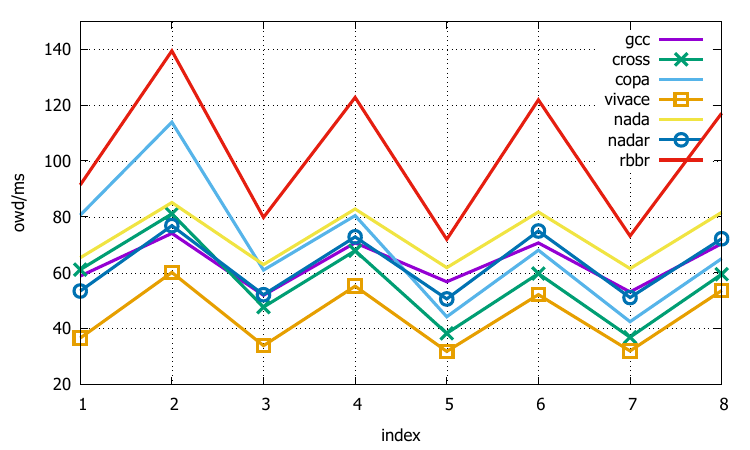}
\caption{Average one-way delay}
\label{Fig:avg-delay}
\end{minipage}
\begin{minipage}[t]{0.49\linewidth}
\centering
\includegraphics[width=1\linewidth]{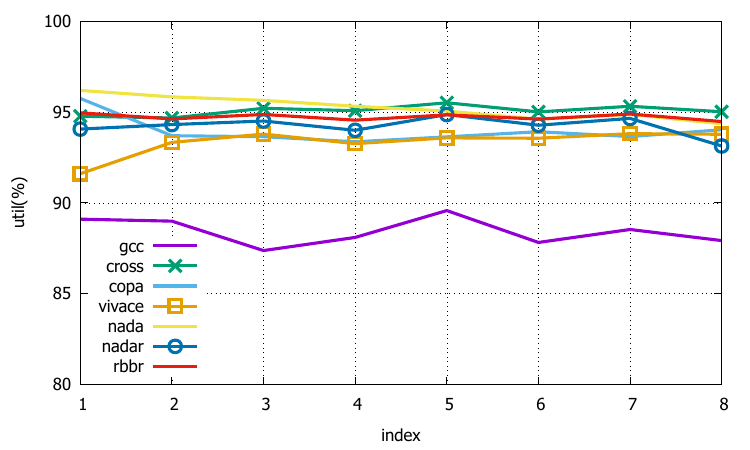}
\caption{Channel utilization}
\label{Fig:channel-util}
\end{minipage}
\end{figure}

\subsection{Intro-protocol Fairness}
Table \ref{tab:link-cfg} contains link configuration. For each experiment, there are three flows and the simulation time is 300 seconds. The first flow runs throughout the entire simulation. The second flow starts at 35 seconds and ends at 200 seconds. The third flow is initiated at 80 seconds and ceases at 300 seconds. All three flows use the same congestion control algorithm for each experiment.

\begin{equation}
\label{eq:util}
util=\frac{\sum_{i}^{n}{sum\_length_{i}}}{C\times T}
\end{equation}
Every 25 milliseconds, the WebRTC sender reads target transfer rate from Call interface and writes it to a trace file. Target transfer rate is updated by congestion controller. When a media packet arrives, the receiver can get packet sent time and sequence number by deserializing the tag object. Hence, one-way transmission delay and packet length can be counted. These delay samples indicate buffer occupation status in the bottleneck. At the end of each session, results on loss rate, average one-way delay and channel utilization are calculated and output to trace files. Equation \ref{eq:util} defines channel utilization. $C$ denotes the capacity of the bottleneck while $T$ is the simulation time. The variable $sum\_length$ represents the total length of received packets and $n$ denotes the number of flows.   

Based on the collected data, the results on average one-way transmission delay and channel utilization are presented in Figure \ref{Fig:avg-delay} and Figure \ref{Fig:channel-util}. GCC has the lowest channel utilization, while RBBR has the highest average transmission delay. When multiple flows compete for resources, the window filter in BBR tends to overestimate the available bandwidth, resulting in bloated delay. Cross dynamically adjusts its rate to maintain a constant queue load in the bottleneck, which has the second lowest transmission delay in most tests. NADAR achieves a lower transmission delay than NADA by periodically cutting rate by half.

\begin{figure}[!htb]
\subfloat[Case4]{
\begin{minipage}[t]{0.5\linewidth}
    \includegraphics[width=1.7in]{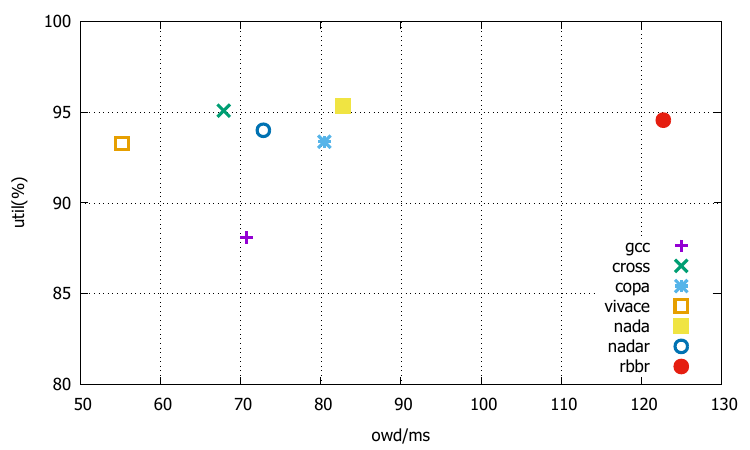}
\end{minipage}}
\subfloat[Case6]{
\begin{minipage}[t]{0.5\linewidth}
    \includegraphics[width=1.7in]{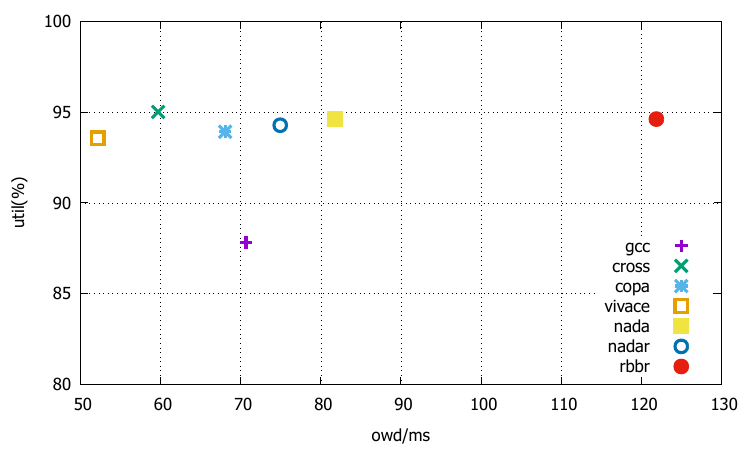}  
\end{minipage}}
\centering  
\caption{Channel utilization and average one-way delay.}
\label{Fig:delay-util-4-6}
\end{figure}

The results on average transmission delay and channel utilization pair in Case4 and Case6 are illustrated in Figure \ref{Fig:delay-util-4-6}. The recently proposed congestion control solutions aim to achieve high throughput and low transmission delay simultaneously. The channel utilization can be seen as the normalized throughput. These algorithms that their working points locate on left top area perform better. Vivace achieves the lowest transmission delay and has an average channel utilization of 93.34\% across all eight tests. However, in a random loss environment, its channel utilization is quite low and Vivace cannot function normally in some tests in the next section. In both figures, Cross achieves similar channel utilization about 95.0\% while maintaining a noticeably lower delay compared to NADA and RBBR.
\begin{figure}[!htb]
\subfloat[Case1]{
\begin{minipage}[t]{0.5\linewidth}
    \includegraphics[width = 1.7in]{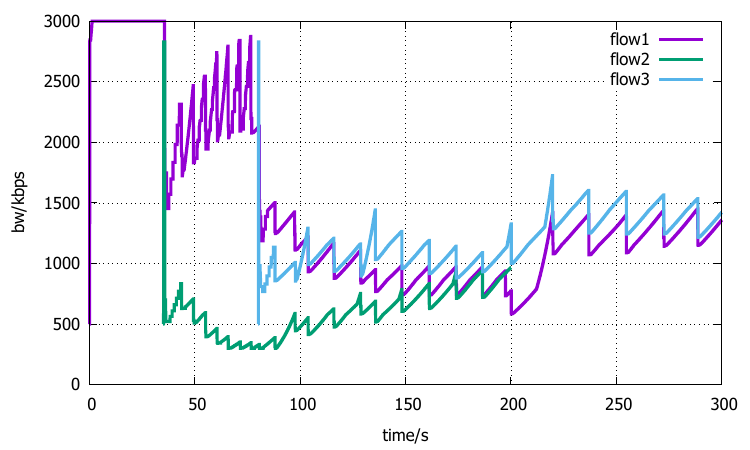}
\end{minipage}}
\subfloat[Case7]{
\begin{minipage}[t]{0.5\linewidth}
    \includegraphics[width = 1.7in]{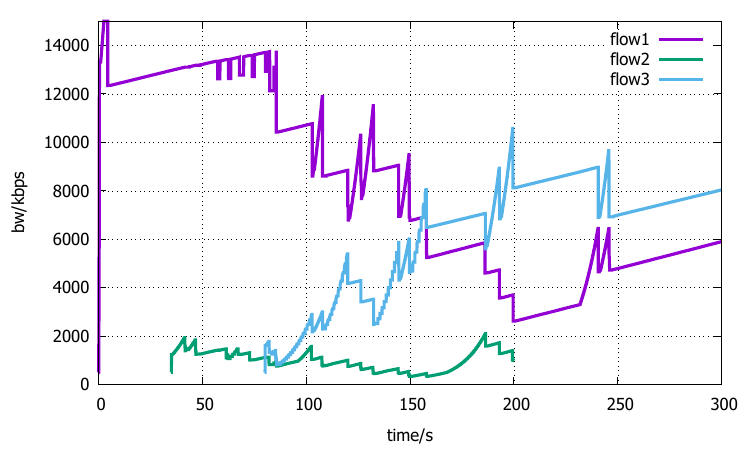}  
\end{minipage}}
\caption{Target transfer rate of GCC flows.}
\label{Fig:gcc-1-5-7} 
\end{figure}
\begin{figure}[!htb]
\subfloat[Case1]{
\begin{minipage}[t]{0.5\linewidth}
    \includegraphics[width = 1.7in]{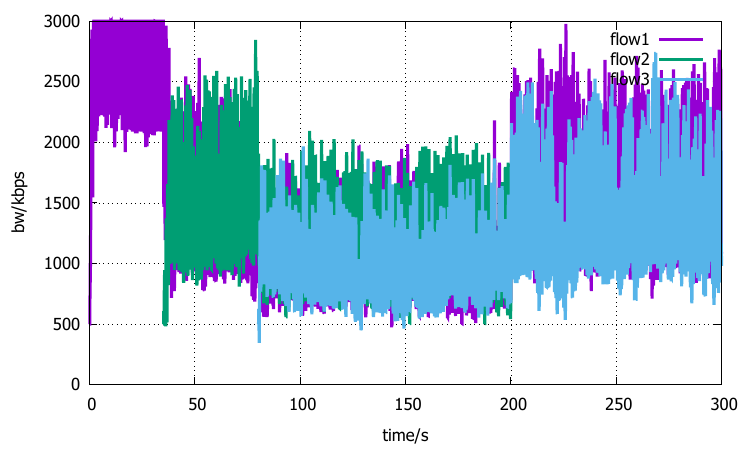}
\end{minipage}}
\subfloat[Case7]{
\begin{minipage}[t]{0.5\linewidth}
    \includegraphics[width = 1.7in]{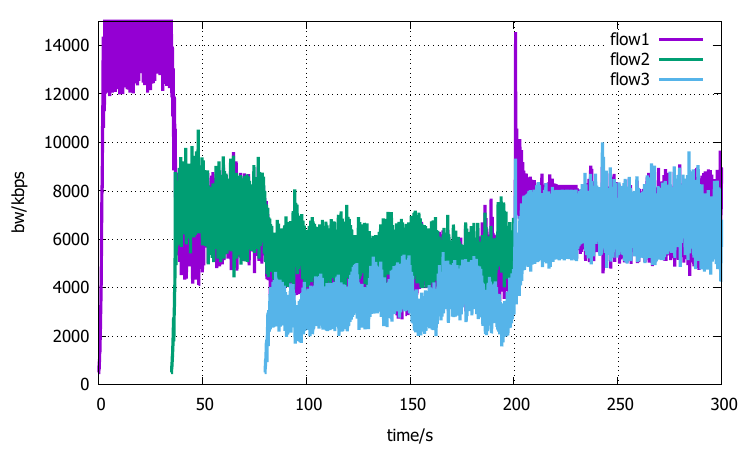}  
\end{minipage}}
\caption{Target transfer rate of Copa flows.}
\label{Fig:copa-1-5-7} 
\end{figure}
\begin{figure}[!htb]
\subfloat[Case1]{
\begin{minipage}[t]{0.5\linewidth}
    \includegraphics[width = 1.7in]{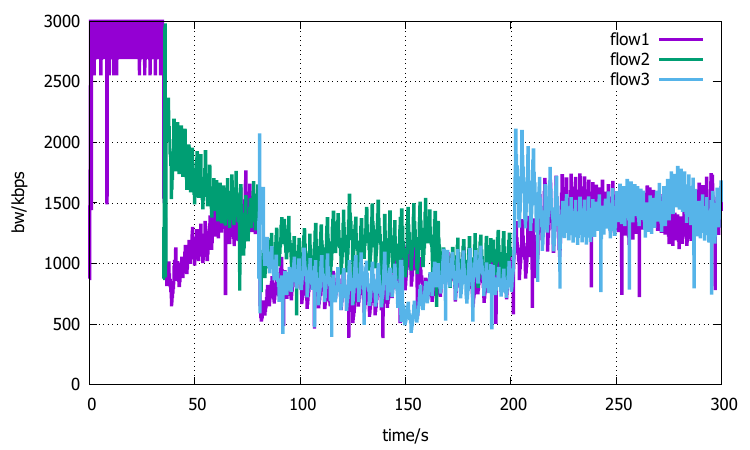}
\end{minipage}}
\subfloat[Case7]{
\begin{minipage}[t]{0.5\linewidth}
    \includegraphics[width = 1.7in]{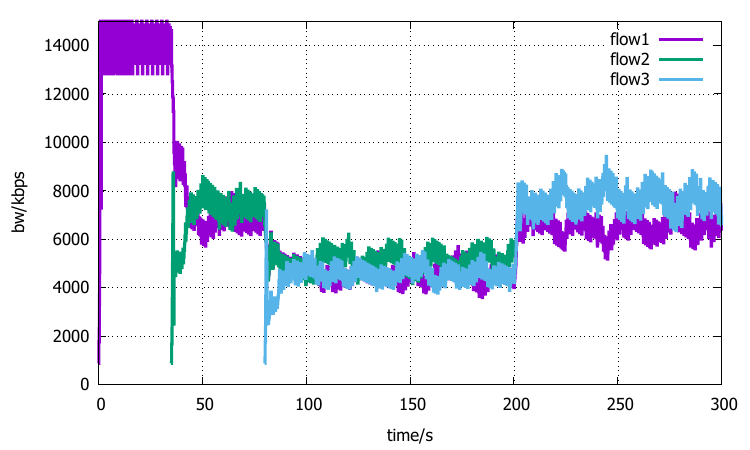}  
\end{minipage}}
\caption{Target transfer rate of Cross flows}
\label{Fig:cross-1-5-7} 
\end{figure}
\begin{figure}[!htb]
\subfloat[Case1]{
\begin{minipage}[t]{0.5\linewidth}
    \includegraphics[width = 1.7in]{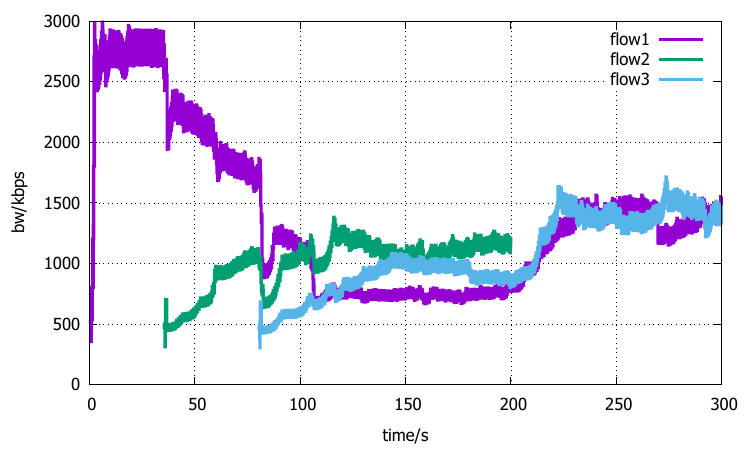}
\end{minipage}}
\subfloat[Case7]{
\begin{minipage}[t]{0.5\linewidth}
    \includegraphics[width = 1.7in]{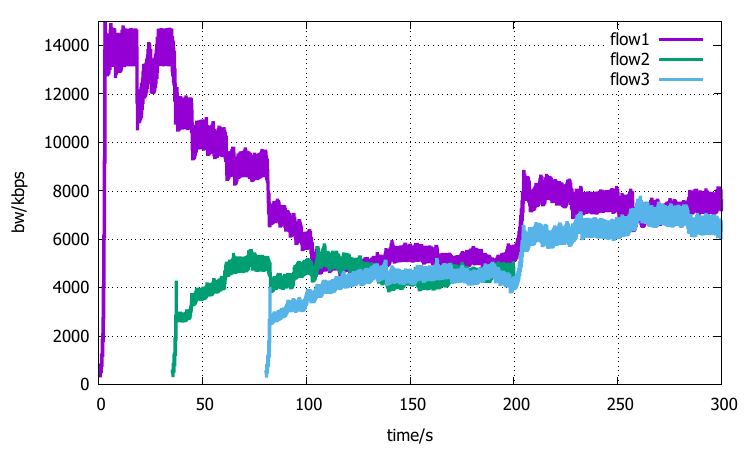}  
\end{minipage}}
\caption{Target transfer rate of Vivace flows.}
\label{Fig:viva-1-5-7} 
\end{figure}

We present the rate dynamic of four algorithms in Case1 and Case7 using collected target transfer rate samples. GCC algorithm is specifically designed for video telephony, a service that requires moderate bandwidth. In Figure \ref{Fig:gcc-1-5-7}(a), the link capacity is 3Mbps and GCC flows can achieve the bandwidth fairness share. However, in a higher-rate link, as shown in Figure \ref{Fig:gcc-1-5-7}(b), the second flow gets starved and its rate is significantly low. These flows do not converge to bandwidth fairness line. Moreover, GCC probes bandwidth slowly. As demonstrated in Figure \ref{Fig:gcc-1-5-7}(b) from 250s, it takes about 50 seconds for flow3 to increase its rate by 1Mbps.

The packet sending rate in Copa is regulated by CWND. Dividing CWND by RTT obtains target tansfer rate. The variation of RTT, the change on CWND and the interval of feedback message lead rate fluctuation as shown in Figure \ref{Fig:copa-1-5-7}. The range of rate amplitude in Cross is smaller when compared with Copa. In Figure \ref{Fig:cross-1-5-7}, the bandwidth allocation fairness in Cross flows can be observed. Cross probes bandwidth with the same gain as Vivace. But  longer time is needed for Vivace flows to converge to throughput in similar level as illustrated in Figure \ref{Fig:viva-1-5-7}(b). For Cross in Figure \ref{Fig:cross-1-5-7}(b), it is quick for flow1 and flow2 to reach rate in similar level. The fast reduce gain in Cross speeds up convergence when new flows come. According to log, the fast reduce gain takes effect after the second flow is initiated.
\begin{figure}[!htb]
\subfloat[NADA]{
\begin{minipage}[t]{0.5\linewidth}
    \includegraphics[width=1.7in]{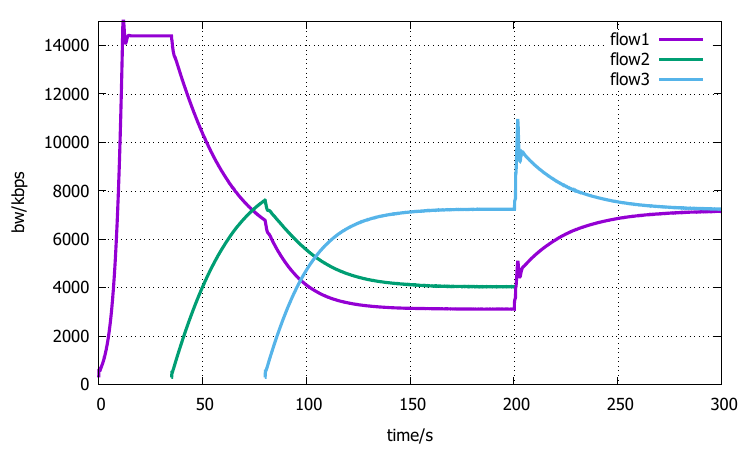}
\end{minipage}}
\subfloat[NADAR]{
\begin{minipage}[t]{0.5\linewidth}
    \includegraphics[width=1.7in]{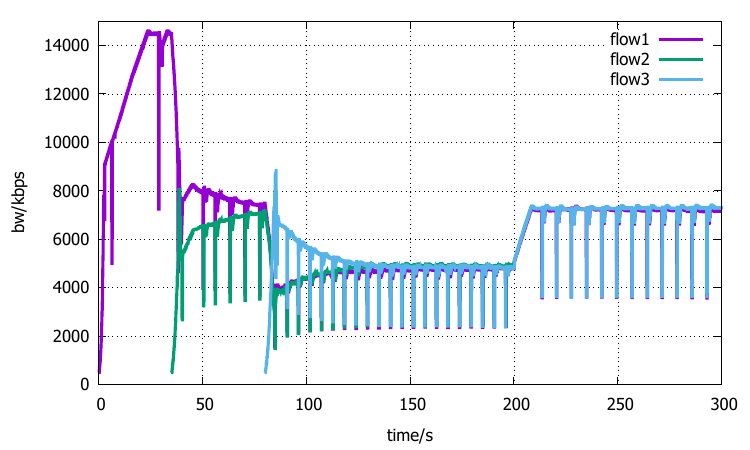}  
\end{minipage}}
\centering  
\caption{Target transfer rate in Case8.}
\label{Fig:rate-nada-nadar-8}
\end{figure}
\begin{figure}[!htb]
\subfloat[NADA]{
\begin{minipage}[t]{0.5\linewidth}
    \includegraphics[width=1.7in]{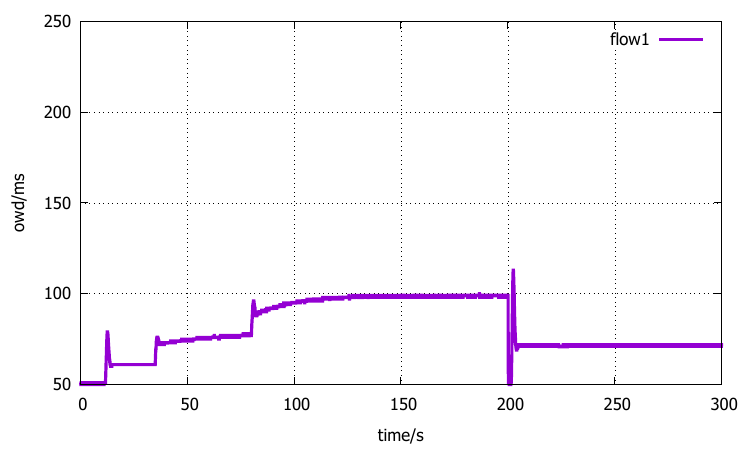}
\end{minipage}}
\subfloat[NADAR]{
\begin{minipage}[t]{0.5\linewidth}
    \includegraphics[width=1.7in]{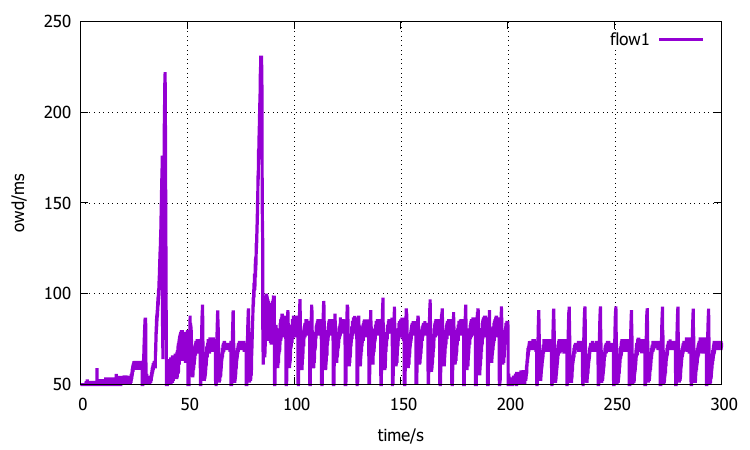}  
\end{minipage}}
\centering  
\caption{One-way delay in Case8.}
\label{Fig:owd-nada-nadar-8}
\end{figure}

As queue load increases, it is the large flows that decrease target rates by rate offset term in NADA. The queue delay is measured as $d_{queue} = d_{fwd}-d_{base}$. Here, $d_{fwd}$ is the measured and filtered one-way transmission delay. $d_{base}$ is the minimal one-way delay during the whole session. The way that NADA controls its rate is similar to a PID (proportional–integral–derivative) controller. NADA overloads the bottleneck and the base delay of new flows is higher than old ones. The congestion extent measured by queue delay for new flows is lesser. NADA suffers the later comer advantage as demonstrated in Figure \ref{Fig:rate-nada-nadar-8}(a) during 100s to 200s. When flow2 exits, the bottleneck has the opportunity to fully drain the buffered packets. A minimal delay appears in Figure \ref{Fig:owd-nada-nadar-8}(a) around time point 200s. Hence, flow3 can sample the same base delay as flow1. Two flows finally converge to a rate in similar level. Experiments in the next section indicate that NADA is sensitive to random packet loss.

ProbeRTT is introduced in NADAR. $d_{fwd}$ is assigned by standing one-way delay, which is obtained from a window filter. When the minimum delay $d_{base}$ is not sampled again within 10 seconds, NADAR enters ProbeRTT and cuts its rate by half. Then, NADAR resamples a new value for $d_{base}$. When multiple flows compete for resource, they all have the chance to sample a minimum value for $d_{base}$ when the link buffer is empty. In Figures \ref{Fig:rate-nada-nadar-8}(b), The ability of NADAR to converge to bandwidth fairness is excellent. 
\subsection{Random Loss Resilience}
A protocol that infers congestion from packet drop events performs poorly in random loss environment. The random packet loss events are not rare in wireless network. Even the average loss rate is quite low in the long run, some sessions can experience instantaneous high loss rate. The seven algorithms are tested in Case 6, where the bottleneck's network device randomly dropped packets with a probability of 0 to 10\% at timepoint 10s. The motivation is to let these loss allergic algorithms to fully probe the maximum available bandwidth at initial stage. 

\begin{figure}
\includegraphics[height=1.5in, width=2.5in]{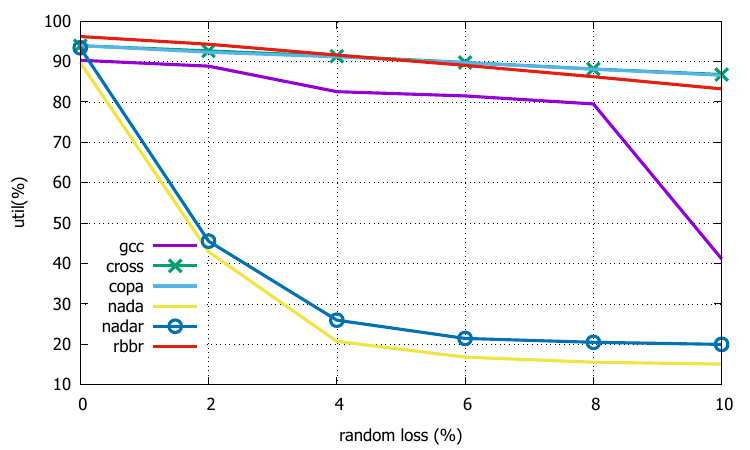}
\centering
\caption{Channel utilization in random loss scenario}
\label{Fig:random-util}
\end{figure}

The results on channel utilization are shown in Figure \ref{Fig:random-util}. When the random packet loss rate is 2\%, the channel utilization of Vivace drops to below 20\%. If the random loss rate is even higher, the simulation process exits abnormally and no meaningful results are obtained. It is possible that there are errors in the implementation of Vivace in WebRTC, which could explain why it has not been applied in practice. 

Unfortunately, the aggregated price to measure congestion in NADA includes a term on packet loss. With a 2\% random loss, NADA's channel utilization is 42.90\%, while NADAR's is 45.52\%. Removing the loss term from the congestion price may improve NADA's channel utilization.

There are two congestion control rules in GCC. One adjusts the rate based on the loss rate and the other regulates the rate based on delay gradient. The target transfer rate is the lowest rate of the two control rules. When the loss rate is above 10\%, the loss-based control rule will aggressively reduce the rate. The channel utilization of GCC is 79.50\% when the loss rate is 8\%. However, when the loss rate increases to 10\%, the channel utilization drops significantly to 41.13\%.

The three algorithms RBBR, Copa and Cross are less affected by random loss. In fact, with 10\% random loss, RBBR has a channel utilization of 83.23\%, while Copa and Cross have a channel utilization of 86.60\% and 86.78\%, respectively.
\subsection{Validation on Fast Bandwidth Adaptation}
A subroutine is incorporated into Cross to enhance its adaptability to scenarios where bandwidth suddenly decreases. Its effectiveness is evaluated in this part. In simulation, the link capacity is 10Mbps at initial, but at 40s, it reduces to 4Mbps before recovering back to 10Mbps at 80s.

\begin{figure}[!htb]
\subfloat[rate dynamic]{
\begin{minipage}[t]{0.5\linewidth}
    \includegraphics[width=1.7in]{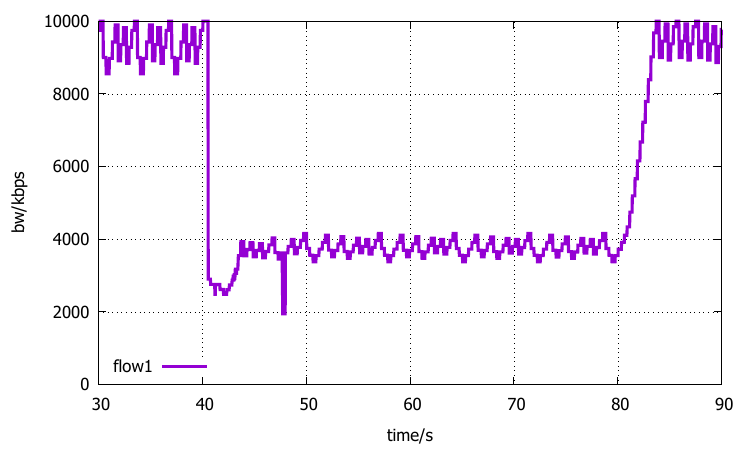}
\end{minipage}}
\subfloat[one-way delay]{
\begin{minipage}[t]{0.5\linewidth}
    \includegraphics[width=1.7in]{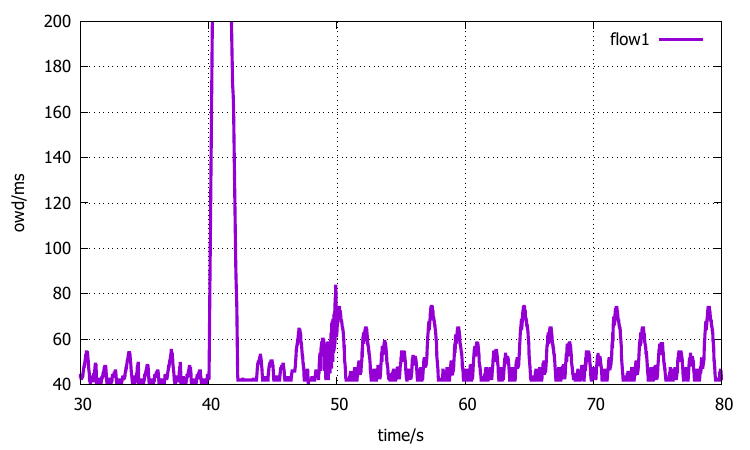}  
\end{minipage}}
\centering  
\caption{Fast bandwidth  adaptation is enabled.}
\label{Fig:cross-change-dec-on}
\end{figure}

\begin{figure}[!htb]
\subfloat[rate dynamic]{
\begin{minipage}[t]{0.5\linewidth}
    \includegraphics[width=1.7in]{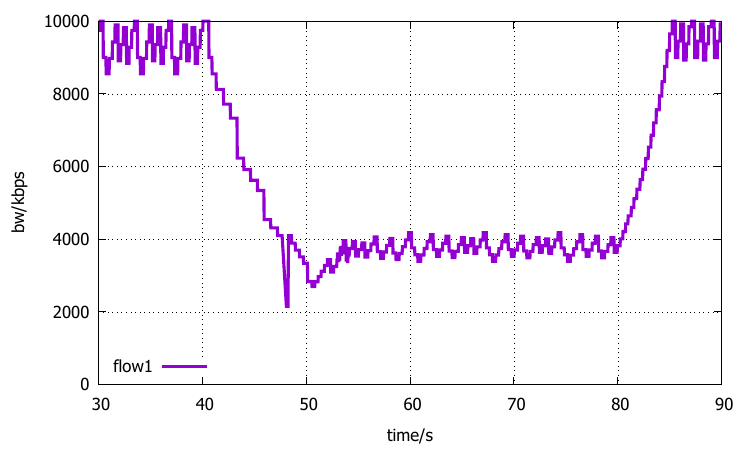}
\end{minipage}}
\subfloat[one-way delay]{
\begin{minipage}[t]{0.5\linewidth}
    \includegraphics[width=1.7in]{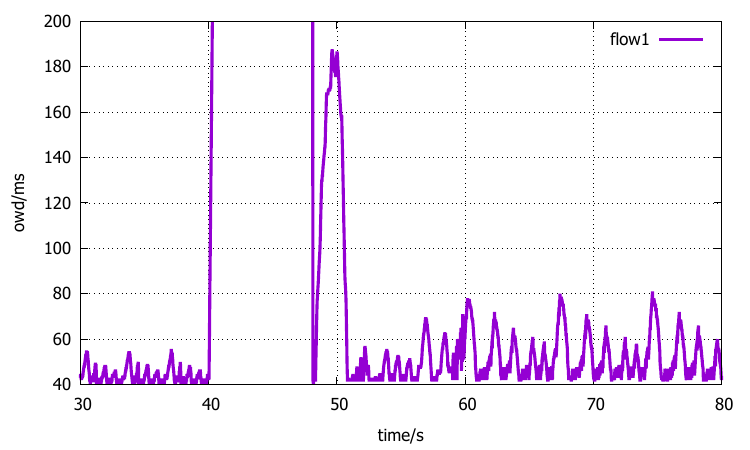}  
\end{minipage}}
\centering  
\caption{Fast bandwidth  adaptation is disabled.}
\label{Fig:cross-change-dec-off}
\end{figure}

When the subroutine is disabled, the results on target transfer rate and one-way transmission delay are illustrated in Figure \ref{Fig:cross-change-dec-off}(a) and (b). As shown in Figure \ref{Fig:cross-change-dec-on}, when the subroutine takes effect, Cross quickly adapts to the dropped capacity. When the bandwidth decreases, the previously sent packets are buffered in the bottleneck and the transmission delay is exceptionally large. In such scenario, Cross resets current bandwidth and re-enters the ProbeRTT state as specified in Algorithm \ref{alg:bw-detection}. By rate reduction at the sender side, the router can drain the buffered packets from the bottleneck. In Figure \ref{Fig:cross-change-dec-on}(a), there is a large latency spike that lasts for approximate 2 seconds, while in Figure \ref{Fig:cross-change-dec-off}(a), the duration of the large latency spike is approximate 10 seconds. Moreover, the switch for fast reduction gain is enabled on in Figure \ref{Fig:cross-change-dec-off}(a).  Without it, the duration of the delay spike would be even longer. The switch for rate acceleration is turned on in Figure \ref{Fig:cross-change-dec-on} (a) and off in Figure \ref{Fig:cross-change-dec-off}. When the capacity recovers at 80s, it takes less time for the flow to return to the peak rate.
\subsection{RTT Fairness}
In an ideal situation, flows employing the same congestion control rule traversing the same bottleneck should achieve throughput at the same level to ensure fairness. However, when flows traverse paths with different propagation delays, many algorithms experience RTT unfairness issues. For example, a Reno flow can gain higher rate when it traverses path with smaller propagation delay while BBR flavors toward flows with larger RTT. 

A dumbbell topology is established, with a bottleneck capacity of 15Mbps. The first path has a roundtrip propagation time of 40ms and flow1 sends packets over it. The roundtrip propagation time of the second path is 80ms and flow2 begins transferring packets over it at 40s. The simulation time is 150 seconds. RBBR, NADAR, and Cross are tested. 

\begin{figure*}[!htb]
\subfloat[RBBR]{
\begin{minipage}[t]{0.3\linewidth}
    \includegraphics[width = 1\linewidth]{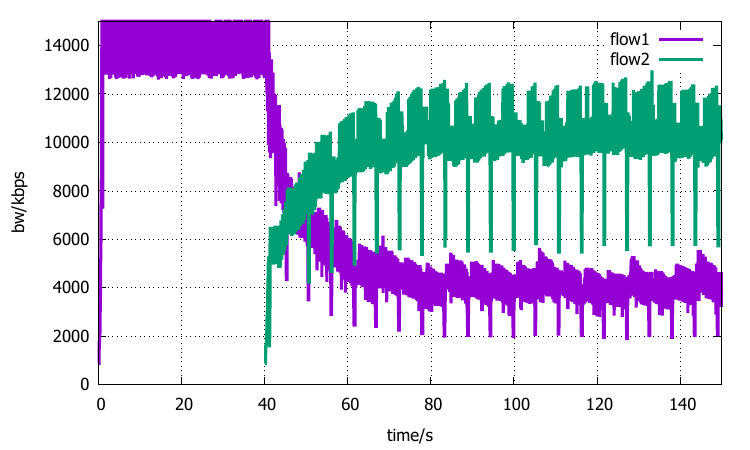}
\end{minipage}}
\subfloat[NADAR]{
\begin{minipage}[t]{0.3\linewidth}
    \includegraphics[width = 1\linewidth]{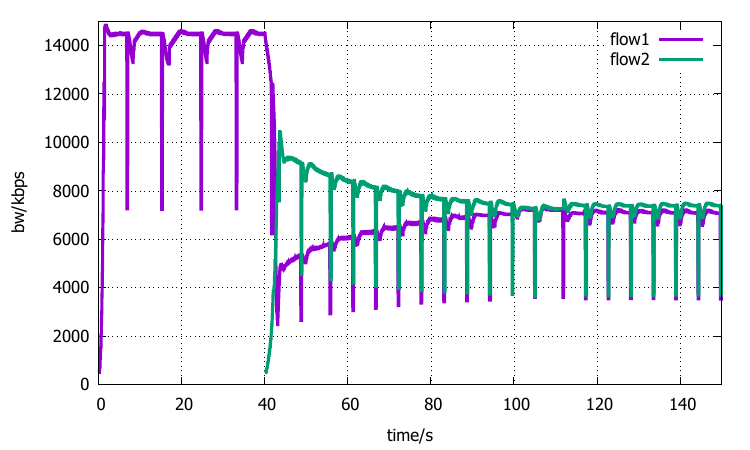}  
\end{minipage}}
\subfloat[Cross]{
\begin{minipage}[t]{0.3\linewidth}
    \includegraphics[width = 1\linewidth]{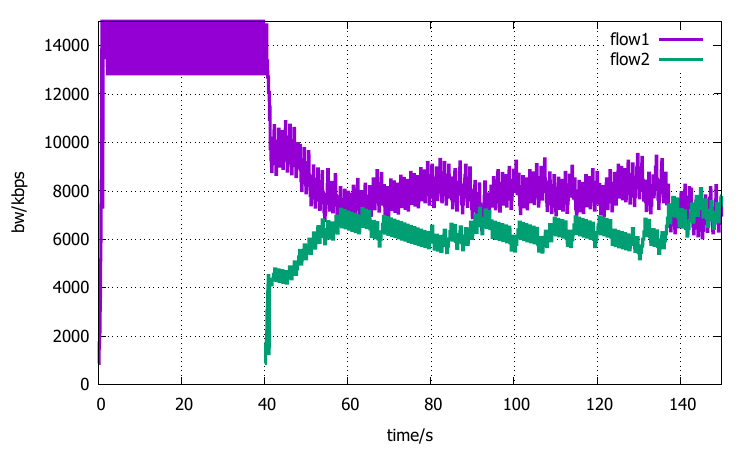}  
\end{minipage}}
\caption{Target transfer rate when flows have different propagation delays.}
\label{Fig:rtt-fair} 
\end{figure*}

The results on target transfer rate are presented in Figure \ref{Fig:rtt-fair}. RBBR experiences the same issue as BBR, with flow2 achieving a higher rate. NADAR and Cross exhibit better performance in term of fairness. The rate gap between two NADAR flows is minimal. Both NADAR and Cross employ queue delay to control rate and penalize flows with a large rate when link falls into congestion. Hence, flows traversing the same bottleneck can finally converge to rate in similar level.

\subsection{Online Evaluation}
We then evaluate the performance of Cross on a virtual museum service, which is similar to cloud gaming. The video frames are rendered by the Unreal engine and then encoded using the H.264 codec. The maximum bitrate of the service is approximate 6Mbps. Before packets reach the browser, we employ traffic control commands to throttle the bandwidth of egress traffic in a Linux PC. It has two Ethernet interfaces and serves as a router. Six different scripts are used to regulate bandwidth. In the first three scripts, the network interface card is configured with a square wave bandwidth. The base bandwidth is set at 3Mbps, 4Mbps, and 6Mbps for Case1, Case2, and Case3, respectively. After that, the bandwidth limit is removed for 30 seconds. As illustrated by Figure 19, there are three troughs, each lasting 30 seconds. For Case4, Case5, and Case6, the bandwidth remains constant at 3Mbps, 4Mbps, and 6Mbps, respectively. During the period of the video streaming, a discount factor on the target transfer rate is used to set the bitrate of the encoder.

Under the same traffic control script, we ran Cross and BBR multiple times. At the browser, the freeze count is read every 30 seconds. It calculates the cumulative duration of video freezes experienced by the receiver. When the time interval between two consecutively rendered frames exceeds a predefined threshold, WebRTC adds the interval to the freeze count. The detailed definition of freeze count can be referred to\footnote{https://www.w3.org/TR/webrtc-stats}. The freezing duration ratio is freezeCount/30. And the monitor board will present the average freezing duration ratio of each session. Figure \ref{Fig:freeze-ratio} presents the error bar on the freezing duration ratio. With the increase of bottleneck bandwidth, the freezing duration ratio of the two algorithms decreases. In all six cases, Cross performs better than BBR. In Case4, the average freezing duration ratio of BBR is 6.11\% and 1.47\% for Cross. In Case6, the average freezing duration ratio of BBR is 1.08\% and 0.91\% for Cross.
\begin{figure}
\includegraphics[height=0.5in, width=3in]{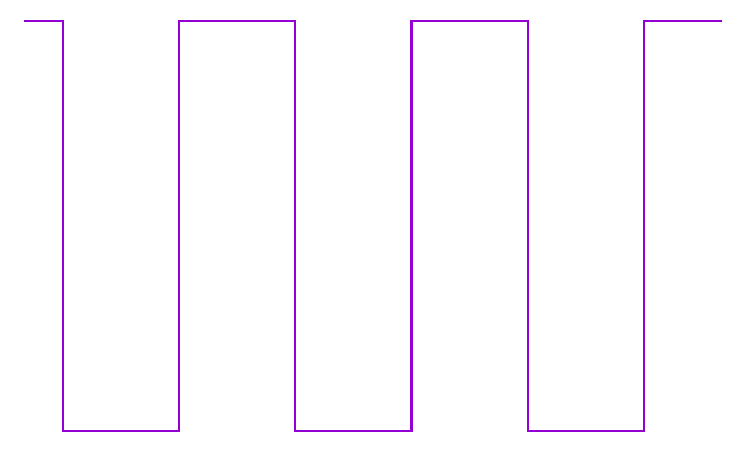}
\centering
\caption{Pictorial on bandwidth throttling}
\label{Fig:squre-wave}
\end{figure}
\begin{figure}
\includegraphics[height=1.5in, width=2.5in]{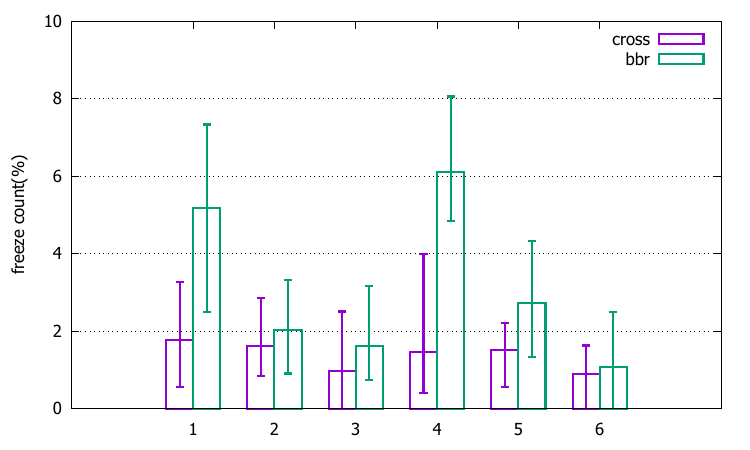}
\centering
\caption{Average video freezing duration ratio}
\label{Fig:freeze-ratio}
\end{figure}
\section{Conclusion}
This work develops a simulation tool that enables WebRTC to be executed on ns-3. By utilizing the simulation module, researchers can develop novel congestion control solutions for WebRTC and evaluate their performance to detect possible anomalies. The module is used to investigate the poor performance of BBR in WebRTC. By introducing two discount factors, RBBR operates normally in WebRTC and reduces packet loss rate significantly. By periodically resampling new values for base delay, the later comer advantage effect issue is solved in NADA. The ability of NADAR to maintain bandwidth allocation fairness is very well but its transmission latency is a bit high. It is worthwhile to further explore the PID-based rate control rule in NADA for real-time applications. By evaluating different algorithms, developers can gain insights into their operating characteristics and choose an appropriate solution for their applications.

Moreover, we introduce a delay-based congestion control algorithm called Cross, which is primarily aimed for high-quality real-time video streaming. Cross adjusts the rate in a MIMD manner by determining whether the queue load exceeds a predefined threshold. This design allows Cross to promptly respond to queue accumulation at bottlenecks. When the bottleneck link experiences congestion, flows with higher rates will yield bandwidth to low rate flows. This behavior promotes fair bandwidth allocation. Simulation results demonstrate that Cross achieves lower delays and maintains considerable channel utilization in networks with random loss. To enhance the quality of experience in networks with significant bandwidth fluctuations, a fast bandwidth adaptation subroutine is integrated into Cross. When such event occurs, the rate is reset according to the newly estimated bandwidth. In the wireless access link, the ability to rapidly adapt to available bandwidth can reduce self-induced queue delays and video freezing.

\section*{Acknowledgment}


%





\ifCLASSOPTIONcaptionsoff
  \newpage
\fi





\bibliographystyle{IEEEtran}
\bibliography{references}
%

\begin{IEEEbiographynophoto}{Songyang Zhang}
received the B.E and Ph.D. degrees in 2015 and 2022 from the College of Computer Science and Engineering, Northeastern University, China. His research interests include real-time video streaming optimization, network congestion control, and multipath transmission.
\end{IEEEbiographynophoto}

\begin{IEEEbiographynophoto}{Changpeng Yang}
received the B.S. degree in Automation Science from Northwestern Polytechnical University, Xi’an, China, the M.S. degree in Pattern Recognition and Intelligent System from Shanghai Jiao Tong University, and Ph.D. in operations research from the Nanyang Technological University, Singapore. He was also affiliated with NEXTOR II in the University of California, Berkeley, USA.  Changpeng Yang is currently director of Huawei Cloud Media Innovation Lab in Huawei Cloud. His research interests include network optimization and decision making in transportation and cloud systems.
\end{IEEEbiographynophoto}






\vfill


\end{document}